\documentclass[aps,prb,showpacs,twocolumn,floats,floatfix]{revtex4}

\usepackage{amssymb}
\usepackage{amsbsy}
\usepackage{amsmath}
\usepackage{epsfig}
\usepackage{graphics}
\usepackage{graphicx}

\newcommand{\bib}{\bibitem}
\newcommand\bea{\begin{eqnarray}}
\newcommand\eea{\end{eqnarray}}
\newcommand\beq{\begin{equation}}
\newcommand\eeq{\end{equation}}
\newcommand\non{\nonumber}

\begin{document}

\title{Infinite-range Ising ferromagnet in a time-dependent transverse 
magnetic field: quench and ac dynamics near the quantum critical point}

\author{Arnab Das$^{(1)}$, K. Sengupta$^{(1)}$, Diptiman Sen$^{(2)}$
and Bikas K. Chakrabarti$^{(1)}$}

\affiliation{$^{(1)}$Theoretical Condensed Matter Physics Division and Center
for Applied Mathematics and Computational Science, Saha Institute of Nuclear
Physics, 1/AF Bidhannagar, Kolkata 700064, India \\
$^{(2)}$Center for High Energy Physics, Indian Institute of Science, Bangalore
560012, India}

\date{\today}

\begin{abstract}
We study an infinite range ferromagnetic Ising model in the presence of a
transverse magnetic field which exhibits a quantum paramagnetic-ferromagnetic
phase transition at a critical value of the transverse field. In the
thermodynamic limit, the low-temperature properties of this model are
dominated by the behavior of a single large classical spin governed by an
anisotropic Hamiltonian. Using this property, we study the quench and AC
dynamics of the model both numerically and analytically, and develop a
correspondence between the classical phase space dynamics of a single spin
and the quantum dynamics of the infinite-range ferromagnetic Ising model. In
particular, we compare the behavior of the equal-time order parameter
correlation function both near to and away from the quantum critical point in
the presence of a quench or AC transverse field. We explicitly demonstrate
that a clear signature of the quantum critical point can be obtained by
studying the AC dynamics of the system even in the classical limit. We
discuss possible realizations of our model in experimental systems.
\end{abstract}


\maketitle

\section{Introduction}

Quantum phase transitions have been studied extensively for several
systems \cite{sachdev}. More often than not, simple prototype
systems such as the Ising model in a transverse field or rotor
models bring out many important characteristics of such transitions
\cite{sachdev,chakrabarti1}. In most of these studies so far, only
the equilibrium situation, where the system is taken adiabatically
across the quantum critical point (QCP), has been addressed
\cite{sachdev,chakrabarti1}. It is well known that in such cases,
the thermodynamic properties of the system can be charted by
determining the critical exponents and hence the universality class
of the transition \cite{sachdev,chakrabarti1,ma}. However, no such general
criteria exists for non-equilibrium dynamics of quantum critical systems.

In recent years, there has been both theoretical and experimental
progress in understanding the behavior of non-equilibrium dynamics near
QCPs. On the experimental side, there have been several experiments
on ultra-cold atoms in optical lattices which realize such
non-equilibrium situations \cite{bloch1}. On the theoretical side,
there have been studies of one-dimensional Ising models in a transverse field
and their variants \cite{sengupta1}, driven quantum spin chains
\cite{cherng1}, and correlation functions for quench dynamics close to a
quantum critical point \cite{calabrese}. The main additional difficulty
over the equilibrium case that one encounters in these theoretical
studies arises from the fact that the description of non-equilibrium
dynamics near a QCP necessitates the knowledge of all or at least
the first few excited states in the Hilbert space of the many-body
Hamiltonian. This precludes a detailed study of most systems which
can be realized experimentally. The notable exceptions are
situations where the systems can be described by either integrable
models \cite{sengupta1, cherng1}, or can be analyzed numerically by exact
diagonalization methods applicable to finite-size systems \cite{sengupta1}.

In this work, we study both the quench and the AC dynamics of an
infinite-range ferromagnetic Ising model in the presence of a
transverse field near its QCP. Such a model, as we show in this
work, can be represented by a single large spin. This particular
feature of the model allows us to analyze its non-equilibrium
dynamics near the QCP. Further, the value of this single spin
increases linearly with the system size so that the large spin and
thermodynamic limits coincide for the model. This enables us to
describe the system in the thermodynamic limit accurately using
classical equation of motions for a single spin. In particular, we
can develop a correspondence between the classical phase space
dynamics of a single spin and the quantum dynamics of the
infinite-range ferromagnetic Ising model in the presence of a
transverse field. In view of the long range dipole-dipole
interactions involved in many systems with order-disorder
transitions driven by tunneling fields, the study of quantum critical
behavior of this model is not just a matter of theoretical curiosity, but
can have application in ferroelectrics \cite{chakrabarti1} like KH$_2$PO$_4$
and ferromagnets like Dy(C$_2$H$_5$SO$_4$)$_3 $9H$_2$O. Further this model
can also can be realized in two-component Bose-Einstein condensates (BEC) of
ultracold atoms \cite{coldref1}.

The main results reported in this work are the following. First, we
obtain the equilibrium phase diagram for the infinite range model
and also obtain its collective excitations using a Holstein-Primakoff
approach for both the ferromagnetic and the paramagnetic phase. Second, we
study the quench dynamics of the equal-time order parameter correlation
function (EOC) and obtain analytical expressions for its long-time
behaviors. Finally, we study the dynamics of the EOC
in the presence of an weak external AC transverse field; we
show that the response of the system in the presence of the AC field
near the QCP involves multiple frequencies and hence appears noisy
in the time domain. This behavior, which persists even in the
classical large $S$ limit of the model, is to be contrasted to that
in both the paramagnetic and the ferromagnetic phases far from the
critical point where the response of the system involves only a few
frequencies and does not exhibit such noisy behavior.
There are some studies on classical Ising models in the presence of
oscillating longitudinal fields \cite{tome-oliv} and some mean field
and Monte Carlo studies \cite{bikas,mjd-oliv} in the presence of an
oscillating transverse field. However, these studies are for completely
different models and none of them investigates the true quantum dynamics.

The organization of the paper is as follows. In Sec. \ref{eqprop},
we introduce the model, chart out its phase diagram and study its
equilibrium properties. In Sec. \ref{qd}, where we
discuss the quench dynamics of the system across the QCP both classically
and quantum mechanically. In Sec. \ref{ac}, we study the AC dynamics of the
model. This is followed in Sec. \ref{summary} by a discussion of possible 
realizations of our model in experimental systems and conclusions.

\section{Equilibrium Properties}
\label{eqprop}

We consider a system of $N$ spin-1/2 objects governed by the Hamiltonian
\beq H ~=~ - ~\frac{J}{N} ~\sum_{i<j} ~S_i^z
S_j^z ~-~ \Gamma ~\sum_i ~S_i^x ~,
\label{ham1} \eeq
where $S^a_i = \hbar \sigma^{a}_i /2$, $a = x, y, z$ are
respectively the $x, y$ and $z$ components of the spin-1/2 operator
represented by the standard Pauli spin matrices $\sigma^a$.
Here we assume that $J \ge 0$ (ferromagnetic Ising interaction). This
Hamiltonian is invariant under the $Z_2$ symmetry $S_i^x \to S_i^x$,
$S_i^y \to - S_i^y$, and $S_i^z \to - S_i^z$.
(The $Z_2$ symmetry would not be present if there was a longitudinal
magnetic field coupling to $\sum_i S_i^z$).
Note that the model in Eq. (\ref{ham1}) differs from the one studied in
Ref. \onlinecite{chakrabarti2}, where the spins were taken to be
living on two sub-lattices, with Ising interactions only between
spins on different sub-lattices. We consider here the ferromagnetic
case, for which the static quantum critical behavior is simple and
easily derivable, while some significant features of its dynamic
critical behaviors are also analytically tractable. We take $\Gamma
\ge 0$ without loss of generality since we can always resort to the
unitary transformation $S_i^x \to - S_i^x$, $S_i^y \to - S_i^y$ and
$S_i^z \to S_i^z$, which flips the sign of $\Gamma$ but leaves $J$
unchanged. Eq. (\ref{ham1}) can be written as
\bea H &=& -~ \frac{J}{2N} ~(~ S_{tot}^z ~)^2 ~-~ \Gamma ~S_{tot}^x ~,
\label{ham2} \\
S_{tot}^z &=& \sum_i ~S_i^z, \quad \quad ~~ S_{tot}^x ~=~ \sum_i ~S_i^x ,
\label{tspin} \eea
and we have dropped a constant $(J/2N) \sum_i (S_i^z)^2 = J/8$ from the
Hamiltonian in Eq. (\ref{ham2}). In the rest of this work, we shall use
units $\hbar =1$.

\subsection{Mean Field Theory}

We begin with a mean field analysis of the thermodynamics of the
model described by Eq. (\ref{ham1}). Denoting the mean field value
$m = \sum_i \left<S_i^z \right>/N$, the Hamiltonian governing any
one of the spins is given by
\beq h ~=~ - J m S^z_{tot} ~-~ \Gamma S^x_{tot} ~.
\eeq
This is a two-state problem whose partition
function can be found at any temperature $T$. If $\beta = 1/k_B T$,
we find that $m$ must satisfy the self-consistent equation
\beq
m ~=~ \frac{Jm}{2 \sqrt{\Gamma^2 + J^2 m^2}} ~\tanh \left( \frac{\beta
\sqrt{ \Gamma^2 + J^2 m^2}}{2} \right) ~.
\label{mag} \eeq
This always has the trivial solution $m=0$. In the limit of zero
temperature, there is a non-trivial solution if $\Gamma < J/2$, with
$|m| = (1/2) \sqrt{1 - 4 \Gamma^2 /J^2}$; the energy gap in that
case is given by $J/2$. If $\Gamma > J/2$, we have $m=0$ and the gap
is given by $\Gamma - J/2$. Hence there is a zero temperature phase
transition at $\Gamma_c = J/2$. The $Z_2$ symmetry mentioned after Eq.
(\ref{ham1}) is spontaneously broken and $<S_i^z>$ becomes non-zero
when one crosses from the paramagnetic
phase at $\Gamma > J/2$ into the ferromagnetic phase $\Gamma < J/2$.

\begin{figure}[htb]
\epsfig{figure=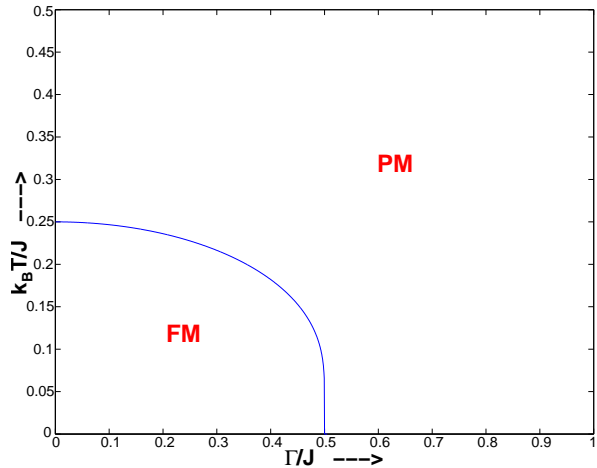,width=8.5cm}
\caption{Phase diagram of the model in mean field theory. FM and PM denote
ferromagnetic and paramagnetic regions respectively.}
\label{figphase}
\end{figure}

In the plane of $(k_B T / J , \Gamma /J)$, there is a ferromagnetic (FM)
region in which the solution with $m \ne 0$ has a lower free energy (the $Z_2$
symmetry is broken), and a paramagnetic (PM) region in which $m=0$. The
boundary between the two is obtained by taking the limit $m \to 0$ in
Eq. (\ref{mag}). This gives $2 \Gamma /J = \tanh (\beta \Gamma /2)$, i.e.,
\beq
\frac{k_B T}{J} ~=~ \frac{\Gamma}{J} ~\left[ \ln \left( \frac{1 + 2
\Gamma /J}{1 - 2 \Gamma /J} \right) \right]^{-1} ~.
\eeq
The phase diagram is shown in Fig. \ref{figphase}.

\subsection{Holstein-Primakoff Approach}

We shall now find the low-energy spectrum of the Hamiltonian in Eq.
(\ref{ham2}). The
form of (\ref{ham2}) shows that the total spin ${\vec S}_{tot}^2 = S(S+1)$ is
a good quantum number; $S$ can take any value from $N/2$ down to 1/2 or 0,
depending on whether $N$ is odd or even. For each value of $S$, the
Hamiltonian describes a single spin; we shall see below that the low
energy properties of the system are dominated by values of $S$ lying close to
$N/2$. The multiplicities $D(S)$ for different values of $S$ (not counting the
trivial multiplicity of $2S+1$ due to different values of $S_{tot}^z$) are
given by the following expressions: $D(N/2) = {}^N C_0 = 1$, $D(N/2-1)=
{}^N C_1 - {}^N C_0 = N-1$, $D(N/2-2) = {}^N C_2 - {}^N C_1 = N(N-3)/2$ and
so on. In general, if $p$ is an integer satisfying $3 \le p \le N/2$, then
\bea & & D(N/2 -p) ~=~ {}^N C_p ~-~ {}^N C_{p-1} \non \\
& & = ~\frac{N(N-1) \cdots (N-p+2)}{p!}~ (N-2p+1) ~.
\label{deg} \eea
For $N \to \infty$ and $p << N/2$, the leading term in $D(N/2 - p)$ is given
by $N^p /p! ~$.

We shall now find the ground state and low-lying excitations of Eq.
(\ref{ham2}) in a sector with a given value of the total spin $S$, assuming
that $\Gamma > J/2$. To do this, we use the Holstein-Primakoff transformation,
\bea S_{tot}^x &=& S ~-~ a^\dagger a ~, \non \\
S_{tot}^z ~-~ i S_{tot}^y &=& \sqrt{2S ~-~ a^\dagger a} ~a ~, \non \\
S_{tot}^z ~+~ i S_{tot}^y &=& a^\dagger ~\sqrt{2S ~-~ a^\dagger a} ~,
\label{hp} \eea
where $[a, a^\dagger] =1$. Assuming that $S$ is close to $N/2$ which is a
large number, we make the approximation of ignoring $a^\dagger a$ with
respect to $2S$ in the last two equations in (\ref{hp}); this gives
\bea S_{tot}^x &=& S ~-~ a^\dagger a ~, \non \\
S_{tot}^z &=& \sqrt{\frac{S}{2}} ~(a ~+~ a^\dagger) ~. \eea
The Hamiltonian in Eq. (\ref{ham2}) then takes the form
\beq
H ~=~ - ~\Gamma S ~+~ \Gamma ~a^\dagger a ~-~ \frac{JS}{4N} ~(a^\dagger ~+~
a)^2 ~. \label{ham3} \eeq
The spectrum of this Hamiltonian can be found by transforming to the
variables $q$ and $p$, where
\beq a ~=~ \frac{q ~+~ ip}{\sqrt 2}~, \quad {\rm and} \quad a^\dagger ~=~
\frac{q ~-~ ip}{\sqrt 2}~, \eeq
and $[q,p]=i$. Then Eq. (\ref{ham3}) takes the form
\beq
H ~=~ - ~\Gamma ~(S +\frac{1}{2}) ~+~ \frac{1}{2} ~\Gamma ~p^2 ~+~ \frac{1}{2}~
(\Gamma ~-~ \frac{JS}{N}) ~q^2 ~. \label{ham4} \eeq
This describes a simple harmonic oscillator with the frequency
\beq \omega_S ~=~ \Gamma ~\sqrt{1 - \frac{JS}{\Gamma N}} ~. \eeq
[Since the maximum value of $S$ is $N/2$, we see that $\omega_S$ is real since
$\Gamma > J/2$. If $\Gamma < J /2$, we have to analyze the problem differently
as discussed below.] The energy spectrum of Eq. (\ref{ham4}) is given by
\beq
E_{S,n} ~=~ - \Gamma ~(S +\frac{1}{2}) ~+~ (n ~+~ \frac{1}{2}) ~\omega_S ~,
\label{en} \eeq
where $n = 0,1,2, \cdots ~$. The ground state of the system corresponds to
$S=N/2$ and $n=0$.

Since $S$ takes the values $N/2, ~N/2 -1, ~N/2 -2, \cdots$, Eq. (\ref{en})
shows that there are two towers of equally spaced excitations: one with a
spacing of $\Gamma$ coming from the first term (arising from excitations
in which the total spin changes) and the other with a spacing $\omega_S$
coming from the second term (arising from excitations in which the total
spin does not change). Only the energy spacing $\Gamma$ turns out
to be thermodynamically significant; this is due to the multiplicities given
in Eq. (\ref{deg}) as we will now see. The partition function of the
oscillator in Eq. (\ref{en}) at an inverse temperature $\beta$ is given by
\beq Z(S) ~=~ \frac{e^{\beta \Gamma (S + 1/2)}}{2 ~\sinh (\beta \omega_S /2)}~.
\eeq
The complete partition function is therefore given by
\beq Z ~=~ \sum_S ~D(S) ~Z(S) ~. \label{part} \eeq
In the thermodynamic limit $N \to \infty$, the sum in Eq. (\ref{part}) will be
dominated by values of $S$ lying close to $N/2$. We can therefore write the
partition function as
\beq Z ~\simeq~ \sum_{p=0}^\infty ~\frac{N^p}{p!} ~\frac{e^{\beta \Gamma
(N/2 - p + 1/2)}}{2 ~\sinh (\beta \omega /2)} ~, \eeq
where we have approximated $\omega$ by its value at $S=N/2$, namely, $\omega
\simeq \Gamma \sqrt{1 - J/2\Gamma}$. The free energy per spin is then given by
\bea f &=& -~ \lim_{N \to \infty} ~\frac{\log Z}{N \beta} \non \\
&=& -~ \frac{\Gamma}{2} ~-~ \frac{e^{-\beta \Gamma}}{\beta} ~, \eea
to lowest order in the quantity $e^{-\beta \Gamma}$, in the limit of zero
temperature. This shows that the thermodynamic gap is given by $\Gamma$.
Note that this agrees with the gap obtained in mean field theory. The
harmonic oscillator energy levels with spacing $\omega_S$ correspond to a
collective excitation of all the spins and therefore they do not appear in
mean field theory (which only takes into account excitations at a single site).

We will end with a brief discussion of the case $\Gamma < J/2$. A
classical analysis with ${\bf S}= \left(S^x, S^y, S_{tot}^z \right)
= S\left( \cos \phi \sin \theta, \sin \phi \sin \theta, \cos \theta
\right)$, where $\theta$ and $\phi$ are the usual polar and
azimuthal angles, gives the Hamiltonian
\beq {\mathcal
H}\left[\theta,\phi \right] ~=~ -~\frac{JS^2}{2N} ~\cos^2 \theta ~-~ \Gamma
S~ \sin \theta \cos \phi ~. \eeq
The lowest energy configuration of this is given by $S=N/2$, $\theta =
\sin^{-1} (2\Gamma /J)$ and $\phi = 0$. There are two solutions for $\theta$,
one lying in the range $[0, \pi /2]$ and the other in the range
$[\pi /2, \pi]$. We choose either one of these as the ground state
and perform a Holstein-Primakoff transformation around it. We then
find that there are again two towers of equally spaced excitations,
one with spacing $J/2$ and the other with spacing $(J/2) \sqrt{1 - 4
\Gamma^2 /J^2}$ corresponding to a collective excitation. The
thermodynamically significant spacing is $J/2$.

Note that as $\Gamma$ approaches $J/2$ from either above or below, the
collective excitation energy softens and goes to zero. Thus it is the
collective excitation gap (not the thermodynamically significant gap) which
is sensitive to the QCP lying at $\Gamma_c = J/2$.

We would like to emphasize here that our model has infinite range interactions
and we are interested in the thermodynamic limit $N \to \infty$. In that limit,
the effective Hamiltonian seen by any one spin is just the mean field produced
by all the other spins. Hence there is no difference between the phase diagrams
obtained by the mean field theory and the Holstein-Primakoff approach.

A difference between mean field theory and the Holstein-Primakoff approach is
that only the latter correctly describes the collective excitation which
involves all the spins (as we have discussed above). However, a single mode
like the collective excitation does not play any role in determining the phase
diagram in the thermodynamic limit.

\section{Quench Dynamics}
\label{qd}

In this section, we will study the quench dynamics of the spin model
across the QCP. We saw above that it is the collective excitations
(in which the total spin does not change) which are sensitive to the
presence of a QCP. We will therefore restrict our attention to the
lowest energy sector in which the total spin is given by $S=N/2$.

A quench dynamics across a QCP has been studied earlier for Ising-like systems
with short-range interactions \cite{sengupta1} and driven XY spin chains
\cite{cherng1}. Here, taking advantage of the fact that the large $S$ (and
hence classical) limit of our model is also the thermodynamic limit, we shall
derive an analytical expression for the long-time average of the EOC in the
thermodynamic limit as the system is quenched across the QCP starting from the
paramagnetic phase. In this section, we shall mostly follow the method of
Ref. \onlinecite{sengupta1}.

\subsection{Quantum Analysis}
\label{qqa}

\begin{figure}[htb]
\epsfig{figure=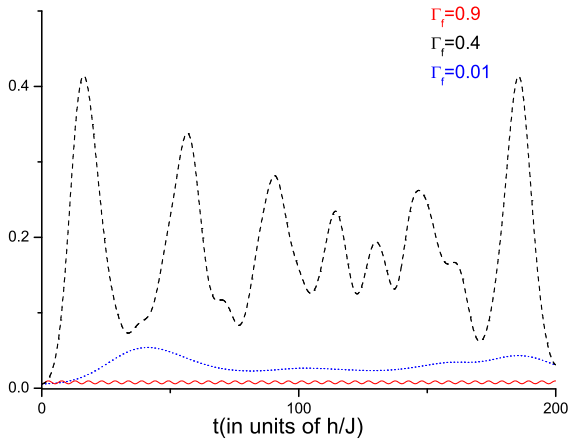,width=8.5cm}
\caption{Dynamics of $\left<(S_{tot}^z)^2\right>/S^2$ for $S=100$ after
quenching the transverse field to different values $\Gamma_f/J$ from an initial
field $\Gamma_i/J=2$. The oscillation amplitudes are small, as seen from the
solid (red) and dotted (blue) curves corresponding to $\Gamma_f/J=0.9$ and
$0.01$ respectively, far away from the critical point $\Gamma_c/J=0.5$. The
oscillation is large in the ordered phase near the critical point as seen from
the dashed (black) curve $\Gamma_f/J=0.4$.}
\label{figdyn1}
\end{figure}

To begin with, we study the dynamics of the EOC (defined as
$\left<(S_{tot}^z)^2\right> /S^2$) by changing the transverse field
$\Gamma$ from an initial value $\Gamma_i \gg \Gamma_c$ to a final
value $\Gamma_f$ suddenly, so that the ground state of the system
has no time to change during the quench. In this case, just after
the quench, the ground state of the system can be expressed, in
terms of the eigenstates $\left|n\right>$ of the new Hamiltonian
${\mathcal H}_f = -(J/4S) (S_{tot}^z)^2 - \Gamma_f S_{tot}^x$ as
\bea \left|\psi \right> &=& \sum_n c_n \left|n\right>, \label{waveini} \eea
where $c_n$ denotes the overlap of the eigenstate $\left|n\right>$
with the old ground state $\left|\psi \right>$. As the state of the
system evolves, it is given at time $t$ by
\bea \left|\psi(t) \right> &=& \sum_n c_n e^{-iE_n t/\hbar}
\left|n\right>, \label{wavedy} \eea
where $E_n = \left<n\right|{\mathcal H}_f \left|n\right>$ are the energy
eigenvalues of the Hamiltonian ${\mathcal H}_f$. The EOC can thus be written as
\bea \left<\psi(t) \right|(S_{tot}^z)^2 /S^2 \left|\psi(t)\right> &=&
\sum_{m,n} c_n c_m \cos \left[\left(E_n-E_m \right)t/\hbar\right] \non \\
&& ~~~~ \times \left< m\right| (S_{tot}^z)^2 /S^2 \left| n \right>.
\label{opcorr} \eea
Eq. (\ref{opcorr}) can be solved numerically to obtain the time
evolution of the EOC. However, certain
qualitative features of the dynamics can be extracted from Eq.
(\ref{opcorr}) without resorting to numerics. First, we note that
when $\Gamma_f$ and $\Gamma_i$ both lie in the paramagnetic phase,
we expect the old ground state to have a large overlap with the new
one, so that $c_n \simeq \delta_{n1}$. In this case,
$\left<(S_{tot}^z)^2(t)\right>$ is expected to undergo small
oscillations about $\left<(S_{tot}^z)^2(t=0)\right>$. On the other
hand, if $\Gamma_f \ll \Gamma_c \ll \Gamma_i$, there is very
little overlap between the two ground states and the amplitude of
oscillation is again expected to be small. Note that the situation
is completely different from the adiabatic turning on of $\Gamma$,
where for $\Gamma_f \ll \Gamma_c$, the system has a maximal value of
$\left<(S_{tot}^z)^2\right>$. In between these two regimes, when
$\Gamma_f \simeq \Gamma_c$, the old ground state is expected to have
significant overlap with many eigenstates $\left|m\right>$, and the
oscillation amplitude can be expected to be large. These qualitative
expectations are verified in Fig. \ref{figdyn1}. Here, we have
quenched the transverse fields to $\Gamma_f/J=0.9,0.01,\,\, {\rm
and}\,\, 0.4$ starting from $\Gamma_i/J=2.0$. The oscillation
amplitudes of the EOC for $S=100$, as shown
in Fig. \ref{figdyn1}, are small for $\Gamma_f=0.9\,\, {\rm and}\,\,
0.01$, whereas it is substantially larger for $\Gamma_f=0.4$.

\begin{figure}[htb]
\epsfig{figure=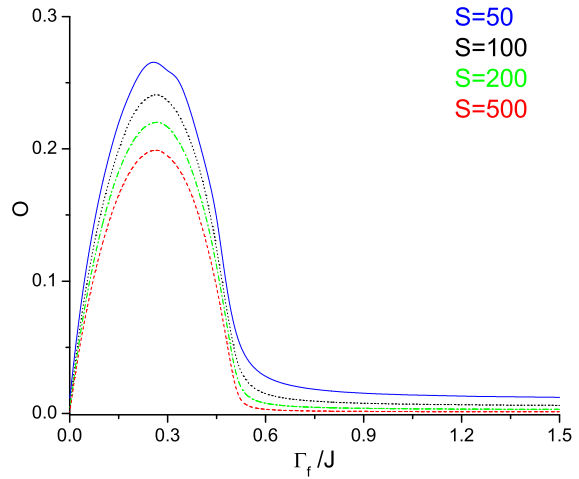,width=8.5cm}
\caption{Plot of the long-time average $O$ as a function of $\Gamma_f/J$ for
different $S$. In the plot the solid (blue), dotted (black), dash-dotted
(green) and the dashed (red) lines represents respectively the results for
$S = 50$, $S = 100$, $S = 200$ and $S = 500$. $O$ peaks around $\Gamma_f/J=
0.25$ and the peak value decreases with increasing $S$. For all plots we have
chosen $\Gamma_i/J=2$.}
\label{figlong}
\end{figure}

Next, to understand the dynamics of the EOC in a little more detail, we study
its long-time averaged value given by
\bea O = \lim_{T\to \infty} \left< \left< (S_{tot}^z)^2(t)\right > \right>_T /
S^2 \non \\
= \frac{1}{S^2} \sum_n c_n^2 \left<n\right|(S_{tot}^z)^2 \left|n \right>
\label{longtime} \eea
for different $\Gamma_f$. Note that the long-time average depends on
the product of the overlap of the state $\left|n\right>$ with the
old ground state and the expectation of $(S_{tot}^z)^2$ in that
state. From our earlier discussion regarding the dynamics of
$\left<(S_{tot}^z)^2 \right>$, we therefore expect $O$ to have a
peak somewhere near the critical point on the ordered side where
such an overlap is maximized. This is verified by explicit numerical
computation of Eq. (\ref{longtime}) in Fig. \ref{figlong} for several values
of $S$ and $\Gamma_i/J=2$. We find that $O$ peaks around $\Gamma_f/J=0.25$,
and the peak height decreases slowly with increasing $S$.

\subsection{Classical Analysis}
\label{qca}

\begin{figure}[htb]
\epsfig{figure=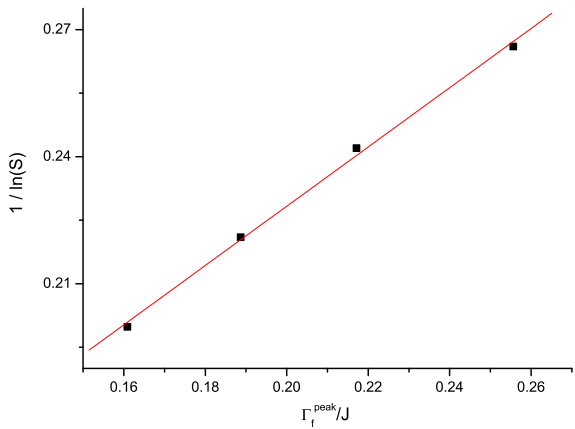,width=8.5cm}
\caption{Plot of the maximum peak height $O_{\rm max}$ of the long-time
average of the EOC as a function of $1/\ln(S)$. The straight line shows the
linear fit.} \label{fig3}
\end{figure}

To understand the position and the system size dependence of the
peak in $O$, we now look at the thermodynamic (large system size)
limit; in the present model, this is also the large $S$ and
therefore classical limit. With this observation, we study the
classical equations of motion for ${\bf S} = S\left(\cos \phi
\sin \theta, \sin \phi \sin \theta, \cos \theta \right)$ for
$\Gamma=\Gamma_f$. In the present model, $S$ is a constant. Thus in
the classical limit, we need to study the equations of motion for
$\theta$ and $\phi$. To this end, we note that the classical
Lagrangian can be written in terms of $\theta$ and $\phi$ as \cite{fradkin}
\bea L &=& - S ~\left[ 1-\cos \theta \right] ~\frac{d \phi}{d t} ~-~ {\mathcal
H}\left[\theta,\phi \right] ~, \label{lagrangian} \eea
This gives the equations of motions
\bea \frac{d \theta}{dt} &=& \Gamma_f ~\sin \phi ~, \non \\
\frac{d \phi}{dt} &=& -~\frac{J}{2} ~\cos \theta ~+~ \Gamma_f ~\cot \theta
\cos \phi ~. \label{eom1} \eea
Eq. (\ref{eom1}) has to be supplemented with the initial condition
that $S_{tot}^x=S$ at $t=0$. The condition $S_{tot}^x=S$ corresponds
to $\theta= \pi/2,\,\,\phi=0$ which is also a fixed point of Eq.
(\ref{eom1}). Therefore we shall start from an initial condition
which is very close to the fixed point: $\theta=\pi/2
-\epsilon, \,\,\phi=\epsilon$, where $\epsilon$ is an arbitrarily
small constant. Further, since the motion occurs on a constant energy
surface after the quench has taken place, we have
\bea
\Gamma_f &=& \frac{J}{4} ~\cos^2 \theta ~+~ \Gamma_f ~ \sin \theta \cos \phi ~.
\label{ec1} \eea
Using Eqs. (\ref{eom1}) and (\ref{ec1}), we get an equation of
motion for $\theta$ in closed form,
\bea \frac{d \theta}{dt} =~ \frac{\sqrt{\Gamma_f^2 \sin^2 \left(\theta\right)
-\left[ \Gamma_f -\frac{J}{4} \cos^2 \theta \right]^2}}{\sin \theta} ~
\equiv f\left(\theta\right). \label{eom2} \eea
It can be seen that the motion of $\theta$ is oscillatory and has
classical turning points at $\theta_1 = \sin^{-1} \left(\left|
1-4\Gamma_f /J \right|\right)$ and $\theta_2 = \pi/2$. One can now
obtain $\left<(S_{tot}^z)^2\right>_T = \left<\cos^2 \theta
\right>_T$ from Eq. (\ref{eom2}),
\bea \left<\cos^2 \theta \right>_T &=& {\mathcal N} /{\mathcal D} ~,
\label{av1} \\
{\mathcal N} &=& \int_{\theta_1}^{\theta_2} d\theta ~\frac{\cos^2
\theta}{f\left(\theta\right)} \non \\
&=& 4 \sqrt{8 \Gamma_f \left(J-2 \Gamma_f \right)}/J ~, \label{num1} \\
{\mathcal D} &=& \int_{\theta_1}^{\theta_2} d\theta ~\frac{1}{f
\left(\theta\right)} ~. \label{deno1} \eea
When trying to evaluate ${\mathcal D}$, we find that the integral
has an end-point singularity at $\theta_2$; this can be regulated
by a cut-off $\eta$ so that $\theta_2 = \pi/2 -\eta$. With this
regularization, ${\mathcal D} = -J \ln(\eta)/\sqrt{\Gamma_f
\left(J-2\Gamma_f\right)/2}$. The cut-off used here has a physical
meaning and is not arbitrary. To see this, note that the angles
$(\theta ,\phi)$ define the surface of a unit sphere of area $4\pi$.
This surface, for a system with spin $S$, is also the phase space which has
$2S+1$ quantum mechanical states. For large $S$, the area of the surface
occupied by each quantum mechanical state is therefore $4\pi /(2S+1) \simeq
2\pi/S$. In other words, each quantum mechanical state will have a linear
dimension of order $1/\sqrt{S}$; this is how close we can get to a given
point on the surface of the sphere. Note that this closeness is determined
purely by quantum fluctuation and vanishes for $S\to \infty$. Thus
$\eta$, which is also a measure of how close to the point $\theta=\pi/2$
we can get, must be of the order of $1/\sqrt{S}$; this
determines the system-size dependence of $\left<\cos^2 \theta \right>_T$.
Using Eq. (\ref{av1}), we finally get
\bea \left<\cos^2 \theta \right>_T &=& \frac{16 \Gamma_f
\left(J -2 \Gamma_f\right)}{J^2 \ln(S)} ~. \label{cr1} \eea
Eq. (\ref{cr1}) is one of the main results of this work. It demonstrates that
the long-time average of the EOC must be peaked at $\Gamma_f/J=0.25$ which
agrees perfectly with the exact quantum mechanical numerical
analysis leading to Fig. \ref{figlong}. Moreover, it provides an
analytical understanding of the $S$ (and hence system-size)
dependence of the peak values of $\Gamma_f/J$. A plot of the peak
height of $O$ as a function of $1/\ln(S)$ indeed
fits a straight line, as shown in Fig. \ref{fig3}. So we conclude
that the peak in $O$ vanishes logarithmically with system size $S$.

\begin{figure}[h]
\centerline{\includegraphics[width=0.7\linewidth,angle=270]{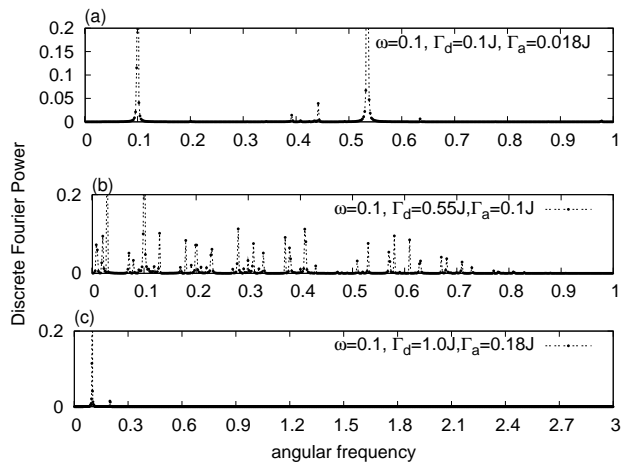}}
\caption{Plot of the DFT power spectrum of the EOC
$\langle\psi(t)| (S_{tot}^z)^2 |\psi(t)\rangle/S^2$ for different
values of $\Gamma_d$. Panels (a) and (c) show the DFTs in the
ferromagnetic and paramagnetic phases respectively, whereas panel
(b) shows the same for the quantum critical region. In all three
panels, the height of the peak at the drive frequency ($\omega =
0.1$) is maximum and is scaled to $1$ (chopped off in the figure).
The plots clearly indicate contributions from a relatively large
number of frequencies to the DFT power spectrum near the QCP.}
\label{ad-AC-fig1}
\end{figure}

\section{AC Dynamics}
\label{ac}

In this section we study the dynamics of the system in the presence of an
oscillatory transverse field. As in Sec. \ref{qd}, we shall first study the
quantum dynamics of the system in the presence of a transverse ac field. This
will be followed by an analysis of the classical equations of motion (Eq.
(\ref{eom1})) which describe the behavior of the system in the large $S$ limit.

\subsection{Quantum Analysis}
\label{aqa}

In the presence of a transverse AC field, the total Hamiltonian is
\bea H(t) &=& -\frac{J}{2N}(S_{tot}^z)^2 - [\Gamma_d + \Gamma_a
\cos (\omega t)]S_{tot}^x, \label{acham} \eea
where $\Gamma_d$ denotes the static part of the transverse field,
$\Gamma_a$ is the amplitude of the AC part, and $\omega$ is the
drive frequency. Note that for small $\Gamma_a$, $\Gamma_d$
determines the point in the phase diagram about which the AC
dynamics takes place. For the rest of this work we shall restrict
ourselves to this limit and for definiteness take a small ratio
$\Gamma_a/\Gamma_d =0.18$. We have checked that the qualitative
picture of the AC dynamics that we describe below remains the same for a
large range of this ratio as long as $\Gamma_a/\Gamma_d \ll 1$.

To elucidate the properties of the AC dynamics of our model, we calculate
the EOC $\langle\psi(t)|(S_{tot}^z)^2 |\psi(t)\rangle/S^2$. For a
sufficiently small driving frequency $\omega$, we are in the
adiabatic regime and the system always stays in the instantaneous
ground state of $H(t)$. In this case, the behavior of the EOC is the
same as that for the static case, and has been well studied
\cite{chakrabarti1}. However, the situation changes when we drive
the system with moderately high frequencies due to possible
Landau-Zener type transitions between different time-dependent
levels of $H(t)$, and a full out-of-equilibrium analysis of the
problem becomes necessary. If, however, the driving frequency becomes very
high compared to other natural frequency scales of the system, the eigenstates
of $H(t)$ vary much faster compared to the characteristic time scale of most
of the possible Landau-Zener transitions. Hence such transitions fail to take
place and the response is characterized by a few trivial frequencies. In this
work, we shall therefore investigate the case of moderately high frequency and
study the characteristics of the frequency spectrum of the EOC. For the sake
of definiteness, we shall choose $\omega=0.1J$ in the rest of this work.

For a given set of parameters ($\Gamma_d, \Gamma_a$ and $\omega$),
we start with the ground state of $H(t=0)$ (in Eq. (\ref{acham})) as
the initial state $|\psi(t=0)\rangle$. We then solve the
time-dependent Schr\"odinger equation to obtain the time
evolution of the state $|\psi(t)\rangle$, using 4th order
Runge-Kutta method with adaptive step-size control \cite{NRF}. The
time variation of the EOC is found, and the discrete Fourier transform
(DFT) of the data is taken to obtain its frequency spectrum.

\begin{figure}[h]
\centerline{\includegraphics[width=0.72\linewidth,angle=270]{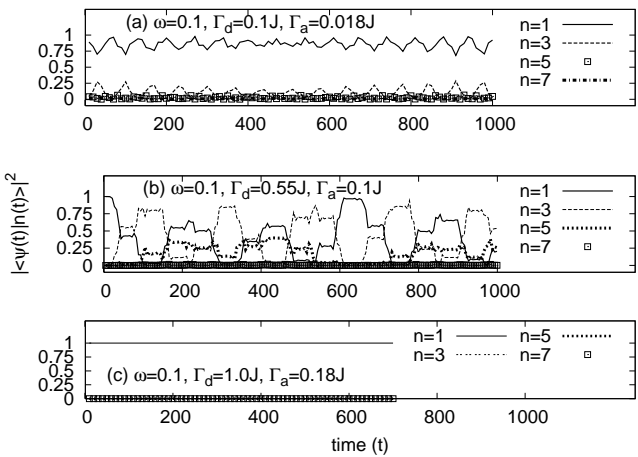}}
\caption{The time evolution (in units of $\hbar/J$) of the overlaps
of the instantaneous state $|\psi(t)\rangle$ with the time-dependent
eigenstates $\left|n(t)\right>$ of the total Hamiltonian $H(t)$ ($n$
= 1 denotes the ground state, $n$ = 2 the first excited state and so
on). Results are shown for $n =$ 1, 3, 5 and 7.}
\label{ad-AC-fig2}
\end{figure}

The DFT power spectra show that inside the paramagnetic and
ferromagnetic phases, the response of the system is dominated by a
few frequencies (panels (a) and (c) of Fig. \ref{ad-AC-fig1}).
These are either the driving frequency $\omega = 0.1J$ or the
natural oscillation frequencies of the system as will be shown in
more details in Sec. \ref{aca}. In contrast, near the QCP (panel (b) of
Fig. \ref{ad-AC-fig1}), several frequencies appear in the spectrum. Thus we
find that the EOC displays a noisy behavior in the time domain near the QPT.

\begin{figure}[h]
\centerline{\includegraphics[width=0.72\linewidth,angle=270]{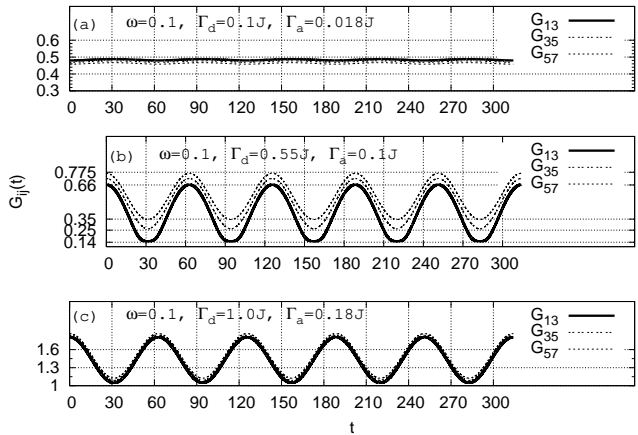}}
\caption{Plot of the relevant gaps $G_{mn}(t) = E_n(t)-E_m(t)$ between
different eigenvalues of $H(t)$ as a function of time (in units of $\hbar/J$)
for different values of $\Gamma_d$. In (a), in the ferromagnetic phase
($\Gamma_d=0.1$), the gaps are almost static and equal at all
times. In (b), near criticality ($\Gamma_d=0.55$),
the gaps vary considerably with time and attain quite low stationary
values. In (c), in the paramagnetic phase ($\Gamma_d=1.0$), the
gaps vary considerably with time, but are always much larger than
their counterparts in the other two phases.} \label{ad-AC-fig3}
\end{figure}

To look into the origin of these frequencies, we have plotted the
instantaneous overlaps of the time dependent state $|\psi(t)\rangle$
with the first few instantaneous eigenstates $|n(t)\rangle$'s of
$H(t)$, for $\omega = 0.1$ in Fig. \ref{ad-AC-fig2}. These overlaps
are significant only for $n = 1$ (ground state), $n = 3, 5$ and $7$
for all times. The overlap is negligible for all other odd $n$ and vanishes
for all even $n$ by symmetry. (The states with odd and even values of $n$
have opposite values of the $Z_2$ parity mentioned after Eq. (\ref{ham1}).
Hence the matrix element of $S_{tot}^x$ between states with odd and even $n$
vanishes). In addition, we have also plotted the instantaneous gaps $G_{mn}(t)
= E_n(t)-E_m(t)$ in Fig. \ref{ad-AC-fig3}. In the ferromagnetic region, as
shown in panel (a) of Fig. \ref{ad-AC-fig2}, the transitions are mainly
between the two levels, $n = 1$ and $3$. However, the overlaps have much
higher average value for $n=1$ than that for $n = 3$. This indicates that
the system mostly stays close to the ground state. Also, as can be
seen from panel (a) of Fig. \ref{ad-AC-fig3}, the relevant gaps
$G_{mn}(t)$ are almost constant, resulting in a periodic behavior
dominated by the driving frequency $\omega = 0.1J$ and a frequency
$0.52J$ which is close to the almost static value of the gaps. Deep
inside the paramagnetic region (panel (c) of Fig. \ref{ad-AC-fig2}),
the system is found to be in the ground state ($|n =1(t)\rangle$)
throughout the evolution; all other overlaps tend to zero. This is
because in this phase, as can be seen from panel (c) of Fig
\ref{ad-AC-fig3}, the instantaneous gaps $G_{mn}(t) = E_n(t)-E_m(t)$
always remain large and prohibits occurrence of any transition from
the ground state to the higher levels. Thus the dynamics is again
governed by a few frequencies and has a simple periodic behavior. In
contrast, near the QCP (panel (b) of Fig. \ref{ad-AC-fig2}) all of
the levels $n = 1, 3$ and $5$ have substantial overlaps with the
instantaneous ground state ($n = 7$ also has a small contribution).
The overlaps are much flatter and more evenly distributed among the
levels $n = 1$ and $3$ on an average. Further, as can be seen from
panel (b) of Fig. \ref{ad-AC-fig3}, the gaps $G_{mn}$ undergo
appreciable oscillations, creating possibilities of Landau-Zener transitions
at multiple frequencies. Thus, quite generally, the AC dynamics near
the QCP involves multiple frequencies and is expected to exhibit
noisy behavior in the time domain.

\begin{figure}[h]
\centerline{\includegraphics[width=0.6 \linewidth,angle=270]{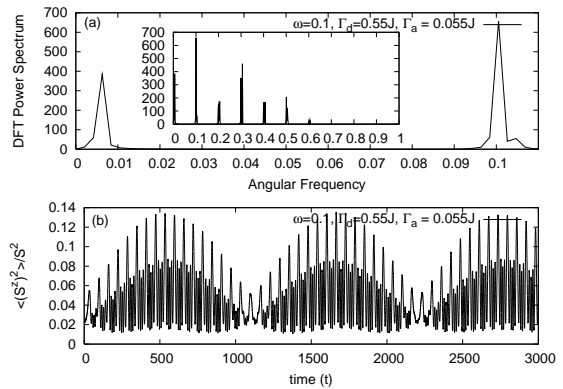}}
\caption{Plot of the DFT power spectrum in panel (a) and the time evolution
in panel (b) of the EOC $\langle\psi(t)| (S_{tot}^z)^2 |\psi(t)\rangle/S^2$
for $S=50$, $\Gamma_d = 0.55$, $\Gamma_a = 0.055$, and $\omega = 0.1$.
Note the oscillations with a very large time period of about 1100 in
panel (b), and the corresponding peak at a frequency of about 0.006 in
panel (a).} \label{rabi}
\end{figure}

\begin{figure}[h]
\centerline{\includegraphics[width=0.6 \linewidth]{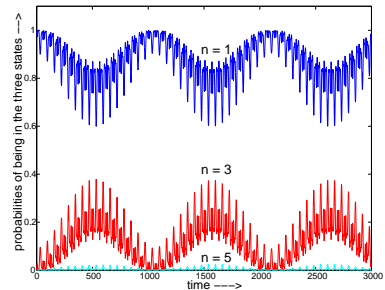}}
\caption{Time evolution of the probabilities of being in different states
when the model considered in Fig. \ref{rabi} is truncated to the three
states $n=$ 1, 3 and 5.} \label{rabi_3state}
\end{figure}

Before ending this section, we would like to point out that the time
evolution of $\left<(S_{tot}^z)^2 \right>$, for some specific
parameter values, may have large oscillations with an unusually long
time period and discuss the reason for its occurrence. To this end,
we present an example of this phenomenon in Fig. \ref{rabi}. The
figure shows oscillations with a time period of about 1100 which is
much longer than the period of the oscillating transverse field, namely,
$2\pi /\omega \simeq 63$. This phenomenon is well-known in the
context of Rabi oscillations which can occur when a two-state system
is subjected to a periodic potential \cite{gas}. For instance, consider a
Hamiltonian which is a $2 \times 2$ matrix of the form $(\omega_0
/2) \sigma^z + (b \cos \omega t) \sigma^x $. If $b$ is small and we
start initially in the ground state of the term $(\omega_0 /2)
\sigma^z$ (the energy gap corresponding to this term is given by
$\omega_0$), we find that if $\omega $ is close to $\omega_0$, the
system periodically makes transitions to the excited state and
returns to the ground state. The time period of this oscillation is
given by $2\pi /\sqrt{(\omega - \omega_0)^2 + b^2}$; this is much
longer than $2 \pi /\omega$ if $\omega /\omega_0$ is close to $1$
and $b \ll \omega$. The amplitude of the oscillations is maximum for
$\omega = \omega_0$. (Similar long time period oscillations can occur
if $\omega /\omega_0$ is close to $1/2, 1/3, \cdots$). Returning to
our problem, for the parameters given in Fig. \ref{rabi}, we find
that the system is quite well described if we only keep the first
three eigenstates corresponding to $n=1,3,5$ of the time-independent
part of the Hamiltonian in Eq. (\ref{acham}). If we truncate the
Hamiltonian to only these three states and study the AC dynamics
starting initially with the first eigenstate, we find that the probabilities
of being in the three states have large and long time period oscillations as
shown in Fig. \ref{rabi_3state}. We thus have a three-state version of
Rabi oscillations. [One of the parameters used in Fig. \ref{rabi}, $\Gamma_a$,
has been chosen to be different compared to the earlier calculations.
This is because the presence of Rabi oscillations is sensitively dependent
on the parameter values; it was necessary to change the value of
$\Gamma_a$, keeping the other parameters fixed, to show a striking
example of such oscillations.] Note that although the probability of
being in the state $n=5$ is small at all times as shown in Fig.
\ref{rabi_3state}, it is necessary to keep this state in order to
obtain the long-time oscillations. We have checked numerically that
if we truncate the Hamiltonian to only the states $n=$ 1 and 3, we
do not get these large and long-time oscillations. We also note that
the driving frequency $\omega = 0.1$ is close to one-fourth of the
gap between the states $n=$ 1 and 3, namely, $G_{13} = 0.408$.
The Rabi oscillations are found to peak at $\omega \simeq 0.101$ and to
disappear if $\omega$ is changed from that value by just $4\%$.
Therefore it is expected that the low-frequency peaks in the DFT spectrum
are not generic features at all drive frequencies, but will be seen
only for special values of the drive frequencies.

\subsection{Classical Analysis}
\label{aca}

The classical analysis of the AC dynamics of the model involves a
numerical solution of Eq. (\ref{eom1}) with $\Gamma_f \to
\Gamma_d + \Gamma_a \cos (\omega t)$. As in the case of the quantum
analysis, we are going to restrict our analysis to an intermediate
drive frequency $\omega=0.1J$ and $\Gamma_d\gg \Gamma_a$. Note that
we are interested in the classical limit of the quantum dynamics,
for which the system is in it's ground state at $t=0$. Thus we
choose the initial condition for the classical dynamics such that
the system is close to its fixed point. These fixed points, as can
be easily seen from Eq. (\ref{eom1}), are given by $\left(\theta_f,
\phi_f\right) = \left(\pi/2,0\right)$ when the system is in the
paramagnetic phase with $\Gamma_d \ge 0.5$, and by $\left(\theta_f,
\phi_f\right) = \left( \sin^{-1} (2\Gamma_d),0\right)$ for the
ferromagnetic phase where $\Gamma_d < 0.5$. Note that the fixed
point structure of Eq. (\ref{eom1}) changes at $\Gamma_d/J=0.5$ which
is also the location of the QCP. Therefore, we choose the initial
condition $(\theta_f-\epsilon, \epsilon)$ for small $\epsilon=0.1$.
We have checked that smaller values of $\epsilon$ do not lead to
qualitative change in the essential features of the dynamics. Also,
in the rest of this section, we shall maintain the same ratio
$\Gamma_a/\Gamma_d=0.18$ as in the previous section.

Before carrying out the numerical simulations, it is useful to
obtain a qualitative understanding of the dynamics of the system in
the presence of the driving field. Since the amplitude of the drive
field $\Gamma_a$ is small and the motion starts from near the
classical fixed point, we expect that a small amplitude analysis for
the motion of $\theta$ and $\phi$ would give us a qualitative
understanding of the dynamics for both the paramagnetic and ferromagnetic
phases. Of course, this expectation is not met near the quantum critical
point, and we are going to discuss this point in more detail subsequently.

\begin{figure}[h]
\hspace*{.0cm}\centerline{\includegraphics[width=0.95\linewidth,angle=270]
{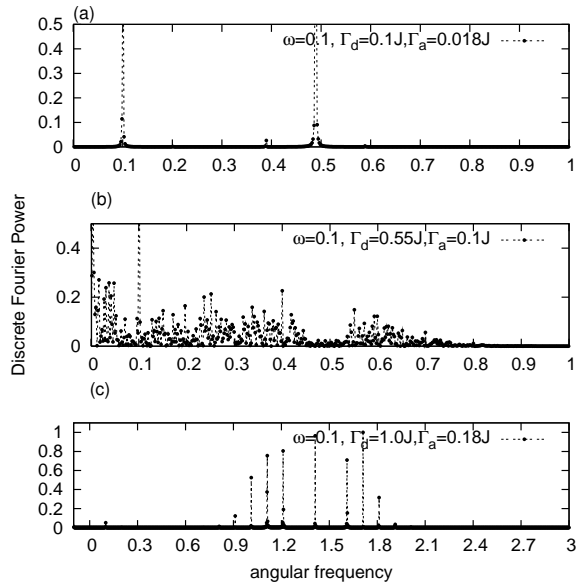}}
\caption{Plot of the DFT power spectrum of the classical EOC $\cos^2 \theta$
for different values of $\Gamma_d$, obtained by solving the classical equations
of motion in Eq. (\ref{eom1}). Panels (a) and (c) show the DFTs in the
ferromagnetic and paramagnetic phases respectively; panel
(b) shows the same for the critical region. Note that the
involvement of many frequencies in the dynamics near the critical
point survives even in the classical limit.} \label{clft}
\end{figure}

Let us first consider the paramagnetic phase where $\Gamma_d > J/2$
and the fixed point is given by $\left(\theta_f, \phi_f\right) =
\left(\pi/2,0\right)$. Linearizing Eqs. (\ref{eom1}) about this fixed
point, one gets a natural frequency of $\omega_n = \sqrt{\Gamma_f
\left(\Gamma_f - J/2 \right)}$ for small amplitude motion. Note
that this analysis is only approximate since for our case $\Gamma_f$
itself varies in time. But for a small drive amplitude and
frequency, where $\Gamma_d - \Gamma_a =0.8J \le \Gamma_f \le
\Gamma_d + \Gamma_a =1.2J$, we expect the main contribution to the
periodic motion of $\theta$ to come from approximately the range
$0.49 J\le \omega_n \le 0.91J$. Thus the power spectrum for the
motion of $(S_{tot}^z)^2(t)/S^2= \cos^2 \theta$, which has half the
time period as $\cos \theta$ in the paramagnetic phase (since $\cos
\theta$ runs over both positive and negative values), is expected to
have significant contribution from frequencies in the range $0.98
J\le \omega_n \le 1.8 J$ and from the drive frequency $\omega =
0.1J$. This expectation can be verified from panel (c) of Fig.
\ref{clft}. The exact location of all the frequencies, however,
cannot be easily deduced from this qualitative argument. Next, let
us consider the ferromagnetic phase at $\Gamma_d=0.1$. Here the
fixed point is given by $\left(\theta_f, \phi_f\right) =
\left(\sin^{-1}(2 \Gamma_f/J),0\right)$. A similar linearization
about the fixed point leads to a natural frequency of $\omega_n =
\sqrt{(J^2/4) - \Gamma_f^2} \simeq 0.5 J$. Note that in this case,
owing to the small ratio of $\Gamma_d/J$, the variation of
$\Gamma_f$ does not affect $\omega_n$, at least to leading order in
$\Gamma_f /J$. Thus we expect that the contribution to the DFT would
arise from the drive frequency $\omega=0.1 J$ and the natural
frequency $\omega_n=0.5 J$. This explains the double peaked
structure of the DFT spectrum in panel (a) of Fig. \ref{clft}.

Finally, let us consider the classical equations at $\Gamma_d=0.55
J$. Here the fixed point at $(\pi/2,0)$ which corresponds to the
energy minima of $H$ (Eq. (\ref{ham2})), is quite shallow and even a
small transverse drive field is enough to drive the system far away from
the fixed point. Therefore we expect the system to traverse a large
part of the phase space during its motion even if $\Gamma_a$ is small.
Consequently, the DFT spectrum of $\cos^2 \theta$ is expected to
have contributions from a wide range of frequencies which translates
to noisy behavior in the time domain. This expectation is verified
by numerical computation as can be seen from panel (b) of Fig. \ref{clft}.
Note that this noisy nature of the AC dynamics of the EOC
occurs due to the presence of a shallow energy minima; this in turn is the
consequence of the presence of a QCP at $\Gamma_d/J=0.5$ where the
fixed point structure of Eq. (\ref{eom1}) changes. Therefore we
expect that such a feature will be generic for QCPs.

\section{Discussion}
\label{summary}

The results obtained in this work can be relevant for certain experimental
systems. One possible class of systems where these results may be applicable
are ones with long range dipole-dipole interactions such as KH$_2$PO$_4$ or
Dy(C$_2$H$_5$SO$_4$)$_3 $9H$_2$O \cite{chakrabarti1} which exhibit
order-disorder transitions driven by tunneling fields. However, the effect of
finite-range of the interactions on the dynamics needs to be studied
carefully before quantitative predictions can be made. We have left
this as a subject of future study.

The other class of systems where our results might be applicable are
two-component BECs where the inter-species interaction ($U$) is
strong compared to the intra-species interaction ($U'$). Such BECs,
in the presence of an external RF drive and a fixed chemical
potential, can be described by an effective Hamiltonian \cite{becth}
\bea H_{\rm eff} &=& \chi S_z^2 + \hbar \Omega S_x \label{bec2} \eea
where the spin operator $S_z = (b_{\uparrow}^{\dagger}
b_{\uparrow}-b_{\downarrow}^{\dagger} b_{\downarrow})/2$ and $S_x =
(b_{\uparrow}^{\dagger}b_{\downarrow}^{\dagger}+
b_{\downarrow}^{\dagger}b_{\uparrow}^{\dagger})/2$ are the $z$ and
$x$ components of the effective spin in terms of the boson operators
$b_{\uparrow,\downarrow}$, $\uparrow$ and $\downarrow$ determines
the species or pseudospin index, $\chi = (U - U')$ is the
coefficient of the $S_z^2$ term, and $\Omega$ is the frequency of
the external RF field. Note that $U'$ can be made arbitrarily strong
compared to $U$ by tuning the system near a Feshbach resonance as
discussed for $^{41}K-^{87}Rb$ system in Ref. \onlinecite{becexp}.
Alternatively, such a Hamiltonian (Eq.\ \ref{bec2}) with $\chi <0$
can also be realized following the method discussed in Refs.\
\onlinecite{becth,bec3}. The quench dynamics described in our work
can then be realized by a sudden change of the frequency of the RF
pulse. A more complicated RF pulse shape can also lead to an
effective sinusoidal time-dependent $\Omega$ whereby the AC dynamics
can also be realized. We also note here there is a vast body of
literature on the model described by Eq. \ref{bec2}
\cite{becth,vidal1}. Most of these works study the entanglement
property of the model for $\chi > 0$. In this regime, the model
(antiferromagnetic infinite-range Ising model in a transverse field)
does not have a second-order quantum phase transition and has
completely different properties than the ferromagnetic model that we
study here.

In conclusion, we have studied the quench and AC dynamics of the EOC
of an infinite range ferromagnetic Ising model, both classically and
quantum mechanically. We have shown that both the quench and the AC
dynamics reflect the presence of the quantum critical point of the
model. In particular, the AC dynamics of the EOC, for a small
amplitude and moderate frequency transverse AC field, exhibits a
noisy behavior near the quantum critical point which is
qualitatively distinct from its periodic behavior away from the
critical point in either the paramagnetic or the ferromagnetic
phase. As we have shown, this noisy behavior of the AC dynamics of
the EOC near the QCP follows from quite general arguments, and
therefore we expect this behavior to qualitatively hold for more
realistic models exhibiting QCP. Finally, we would like to point out
that the results presented here may be relevant for Ising systems
with long range dipolar interactions and two-component BECs.

D.S. thanks DST, India for financial support under the project SP/S2/M-11/2000.
The authors thank J. Vidal for useful correspondence.


\begin{thebibliography}{99}

\bib{sachdev} S. Sachdev, {\it Quantum Phase Transitions} (Cambridge
University Press, Cambridge, 1999).

\bib{chakrabarti1} B. K. Chakrabarti, A. Dutta, and P. Sen, {\it Quantum Ising
Phases and Transitions in Transverse Ising Models} (Springer, Heidelberg,
1996).

\bib{ma} S. K. Ma {\it Modern Theory of Critical Phenomena} (Addison-Wesley,
New York, 1996).

\bib{bloch1} For a review see, I. Bloch, Nature Physics {\bf 1}, 23 (2005).

\bib{sengupta1} K. Sengupta, S. Powell, and S. Sachdev, Phys. Rev. A {\bf 69},
053616 (2004).

\bib{cherng1} R. W. Cherng and L. S. Levitov, Phys. Rev. A {\bf 73}, 043614
(2006).

\bib{calabrese} P. Calabrese and J. Cardy, Phys. Rev. Lett. {\bf 96}, 136801
(2006).

\bib{coldref1} D. Gordon and C. M. Savage, Phys. Rev. A {\bf 59}, 4623 (1999).

\bib{tome-oliv} T. Tom\'{e} and M. J. de Oliveira, Phys. Rev A {\bf 41}, 4251
(1990); A. Krawiecki, Int. J. Mod. Phys. B {\bf 19}, 4769 (2005); G. M.
Buend\'{\i}a and E. Machado, Phys. Rev. E {\bf 58}, 1260 (1988); M. Keskin,
O. Canko, and U. Temizer, Phys. Rev. E {\bf 72}, 036125 (2005).

\bib{bikas} M. Acharyya, B. K. Chakrabarti, and R. B. Stinchcombe, J. Phys. A
{\bf 27} 1533 (1994); V. Banerjee, S. Dattagupta, and P. Sen, Phys. Rev. E
{\bf 52} 1436 (1995).

\bib{mjd-oliv} M. Santos and M. J. de Oliveira, Int. J. Mod. Phys. B {\bf 13},
207 {1999}.

\bib{chakrabarti2} B. K. Chakrabarti and J.-I. Inoue, Ind. J. Phys. {\bf 80},
609 (2006); B. K. Chakrabarti, A. Das, and J.-I. Inoue, Euro. Phys. J. B 
{\bf 51}, 321 (2006).

\bib{fradkin} E. Fradkin, {\it Field Theories of Condensed Matter Systems},
Addison-Wesley, Reading, (1991).

\bib{NRF} W. H. Press, S. A. Teukolsky, W. T. Vetterling, and B. P. Flannery,
{\it Numerical Recipes in Fortran 77: The Art of Scientific Computing}
(2nd Ed.) (Cambridge University Press, Cambridge, 2001).

\bib{gas} S. Gasiorowicz, {\it Quantum Physics} (John Wiley \& Sons, Singapore,
2000).

\bib{becth} A. Micheli, D. Jaksch, J.I. Cirac, and P. Zoller, Phys. Rev A
{\bf 67}, 013607 (2003); A. P. Hines, R. H. McKenzie, and G. J. Milburn, Phys.
Rev A {\bf 67}, 013609 (2004).

\bibitem{bec3} J. Cirac, M. Lewenstein, K. Molmer and P. Zoller, Phys. Rev A 
{\bf 57}, 1208 (1998).

\bib{becexp} A. Simoni, F. Ferlaino, G. Roati, G. Modugno, and M. Inguscio,
Phys. Rev. Lett {\bf 90}, 163202 (2003).

\bib{vidal1} S. Dusuel and J. Vidal, Phys. Rev. Lett. {\bf 93}, 237204 (2004);
{\it ibid}, Phys. Rev. B {\bf 71}, 224420 (2005); J. Vidal, G. Palacios, and
J. Aslangul, Phys. Rev. A {\bf 70}, 062304 (2004).

\end{thebibliography}
\end{document}